\documentclass[aps,prl,twocolumn,showpacs,amsmath,amssymb,groupedaddress]{revtex4}

\usepackage{graphicx}
\usepackage{dcolumn}
\usepackage{bm}

\begin{document}

\preprint{APS/123-QED}

\title{Observation of spinor dynamics in optically trapped $^{87}$Rb
Bose-Einstein Condensates}

\author{M.-S. Chang}
\author{C. D. Hamley}
\author{M. D. Barrett}
\altaffiliation[Current address ]{NIST, Boulder, CO.}
\author{J. A. Sauer}
\author{K. M. Fortier}
\author{W. Zhang}
\author{L. You}
\author{M. S. Chapman}
\affiliation{School of Physics, Georgia Institute of Technology, \\
Atlanta, Georgia 30332-0430}
\date{\today}

\begin{abstract}
We measure spin mixing of F=1 and F=2 spinor condensates of
$^{87}$Rb atoms confined in an optical trap. We determine the spin
mixing time to be typically less than 600 ms and observe spin
population oscillations. The equilibrium spin configuration in the
F=1 manifold is measured for different magnetic fields and found
to show ferromagnetic behavior for low field gradients. An F=2
condensate is created by microwave excitation from F=1 manifold,
and this spin-2 condensate is observed to decay exponentially with
time constant 250 ms. Despite the short lifetime in the F=2
manifold, spin mixing of the condensate is observed within 50 ms.
\end{abstract}

\pacs{03.75.Mn, 32.80.Pj, 03.67.-a}
\maketitle

One of the hallmarks of Bose-Einstein condensation (BEC) in dilute
atomic gases is the relatively weak and well-characterized
inter-atomic interactions that allow quantitative comparison with
theory.  The vast majority of experimental work has involved
single component systems, using magnetic traps confining just one
Zeeman sub-level in the ground state hyperfine manifold. An
important frontier in BEC research is the extension to
multi-component systems, which provides a unique opportunity for
exploring coupled, interacting quantum fluids. In particular,
atomic BECs with internal spin degrees of freedom offer a new form
of coherent matter with complex internal quantum structures. The
first two-component condensate was produced utilizing two
hyperfine states in $^{87}$Rb, and remarkable phenomena such as
phase separation were observed \cite{Wieman971,Wieman98}. Sodium
F=1 spinor BECs have been created by transferring spin polarized
condensates into a far-off resonant optical trap to liberate the
internal spin degrees of freedom. This allowed investigations of
the ground state properties of  Na spinor condensates, and
observations of domain structures, metastability, and quantum spin
tunneling \cite{Ketterle982,Ketterle991,Ketterle992}.

In this letter, we explore the spin dynamics and ground state
properties of $^{87}$Rb spinor condensates in an all-optical trap,
by starting with well-characterized initial conditions in a known
magnetic field. We focus on the F=1 case and confirm the predicted
ferromagnetic behavior.  We observe population oscillation between
different spin states during the spin mixing and observe reduced
magnetization fluctuations, pointing the way to future exploration
of the underlying spin squeezing and spin entanglement predicted
for the system \cite{You032}. We also create F=2 spinors using a
microwave excitation, measure a decay of the condensate with a
time constant of 250 ms. Despite the short lifetime, spin mixing
of the spin-2 condensates is observed within 50 ms.  Similar
results are concurrently reported in Ref \cite{Sengstock03}; in
that work, the emphasis is on the F=2 mixing, while here, we focus
mainly on the F=1 manifold.

A spinor BEC can be described by a multi-component order parameter
which is invariant under gauge transformation and rotation in spin
space \cite{Ho98,MachidaOhmi98,Bigelow98}. For a spin-1 BEC, the
condensate is either ferromagnetic or anti-ferromagnetic
\cite{Ho98}, and the corresponding ground state structure and
dynamical properties of these two cases are very distinct. The Na
F=1 spinor was found to be anti-ferromagnetic, while the F=1
$^{87}$Rb is predicted to be ferromagnetic
\cite{Greene01,Verhaar02}. Even richer dynamics are predicted for
spin-2 condensates \cite{Ho00}, although they remain largely
unexplored experimentally \cite{Sengstock03,Ketterle031}.

Single component BECs are typically well-described by a scalar
order parameter $\psi (\mathbf{r},t)$ (the BEC ``wavefunction")
whose dynamics is governed by the nonlinear Gross-Pitaevskii
equation \cite{Pethick02}, $i\hbar \frac{\partial \psi}{\partial
t} = -\frac{\hbar ^2}{2m} \triangledown ^2 \psi +V_{t}\psi
+c_{0}|\psi |^2\psi$, where $V_{t}$ is the trap potential, $c_{0}$
the two-body mean field interaction coefficient, and $|\psi |^2=n$
is the density. For spinor condensates, the Bogoliubov formalism
can be extended to a vector order parameter $\psi \mathbf{\chi}$,
where $\mathbf{\chi}$ is the spin vector with 2F+1 components for
spin-F condensates\cite{Ho98,MachidaOhmi98}. For F=1, the two-body
interaction energy including spin  is
$U(r)=\delta(r)(c_{0}+c_{2}\mathbf{F_{1}} \cdot \mathbf{F_{2}})$,
where $r$ is the distance between two atoms and $c_{2}$ is the
spin dependent mean-field interaction coefficient. For F=2,
$U(r)=\delta(r)(\alpha+\beta\mathbf{F_{1}} \cdot
\mathbf{F_{2}}+5\gamma P_{0})$, where $\alpha$ is a
spin-independent coefficient, $\beta$ and $\gamma$ are
spin-dependent coefficients, and $P_{0}$ is the projection
operator \cite{Ho98,MachidaOhmi98}.

The spin-1 ground states have been calculated by several groups
for both zero magnetic field \cite{Ho98, Bigelow98} and finite
field cases \cite{You031}. Although the spin dependent terms, e.g.
$|c_{2}|$, are typically one to two orders of magnitude less than
the spin independent mean field interaction $|c_{0}|$, they can
have a dramatic effect on the ground state structure and dynamical
properties of the condensates.  Ho showed that if $c_{2}$ is
negative (positive), the spinor displays ferromagnetic
(anti-ferromagnetic) behavior \cite{Ho98}. For the case of F=1
$^{87}$Rb, $c_{2}$ is calculated to be $-3.58(57)\times
10^{-20}\mathrm{Hz\cdot m^{3}}$ \cite{Verhaar02}. Hence the ground
state of $^{87}$Rb should be ferromagnetic \cite{Greene01} with a
global minimal energy state at zero field of $|F=1,m_{F}=1\rangle$
\cite{Ho98}.

For finite fields, the ferromagnetism will manifest itself in
different ways depending on the scale of the ferromagnetic energy
compared with the Zeeman energy.  For a typical density of
$n=4\times 10^{14}$ cm$^{-3}$ in our optical traps, the
ferromagnetic energy $2|c_{2}|n$ is only 28 Hz \cite{Ketterle982},
and hence, observation of the low field ground state,
$|F=1,m_{F}=1\rangle$, requires that the first order (linear)
Zeeman shift $E_{Z}=m_{F}B\times (700$ Hz/mG) be smaller than
$2|c_{2}|n$, and hence requires zeroing the magnetic field $B$ to
less than 40 $\mu$G, typically requiring a magnetically shielded
environment.

Even at much larger fields (up to $\sim$500 mG in our case),
however, the ferromagnetic interactions can still play a dominant
role in determining the spin ground state  due to constraints
imposed by angular momentum conservation.  If we start with a
non-equilibrium spin mixture, the system will relax to the minimal
energy state via spin exchange collisions, and for the F=1
manifold, the only processes that conserve angular momentum are
\begin{equation}\label{eqn:spin-mix}
2|m_{0}\rangle \rightleftharpoons |m_{1}\rangle + |m_{-1}\rangle.
\end{equation}

Hence, the normalized magnetization of the system $M=n_{1}-n_{-1}$
is a conserved quantity, where $n_{i}$ is the fraction of
condensate atoms in the $i$th Zeeman state $m_{i}$, and for a
given $M$, the system is uniquely determined by $n_{0}$. In this
field regime, the effect of the anti-ferromagnetism
(ferromagnetism) is to lower (raise) the energy of the $m_{0}$
spinor component relative to the average energy of the $m_{\pm 1}$
components \cite{Ketterle982}, which drives the reaction in Eq.\
\ref{eqn:spin-mix} to the left (right). Hence, the extent to which
this reaction is driven provides an unambiguous distinction
between the ferro- and anti-ferromagnetic cases.  At higher
fields, these effects compete with the second order (quadratic)
Zeeman shift, $(E_{+1}+E_{-1}-2E_{0})/2\simeq B^{2}\times$ (350
Hz/G$^{2}$) \cite{Ketterle982}, which tends to drive the reaction
to the left (to the $m_{0}$ state) for fields $B\gtrsim$ 500 mG.
Although evidence of the ferromagnetism of $^{87}$Rb was already
provided by the observed spin mixtures in our previous work (which
were measured at $\sim$100 mG) \cite{Chapman01, Chapman02}, and
currently reported by Ref. \cite{Sengstock03} (which are measured
at $\sim$340 mG), here we present systematic studies of the spin
mixing, starting with non-equilibrium initial conditions and
following its time evolution for different magnetic fields.

Our experiments employ an all-optical BEC technique previously
described in \cite{Chapman01}.  The atoms are loaded directly from
a magneto-optical trap into an optical dipole force trap formed by
a CO$_2$ laser. Lowering the laser power forces evaporative
cooling in the optical trap, leading to rapid condensation. In
this work, we employ a large period (5.3 $\mu$m) 1-D lattice made
by a CO$_{2}$ laser standing wave \cite{Chapman02} which provides
a strongly anisotropic pancake shape trap, allowing clear
distinction between thermal clouds and condensates. The lattice is
loaded by transferring atoms from an orthogonal travelling wave
trap. We create condensates in only one lattice site by adjusting
the trap power during transfer. The condensates contain $30,000$
atoms in a single lattice site with measured trap frequencies
$2\pi(120, 120, 2550)$ rad/s, and no observable thermal component.
The density in the optical trap is estimated to be
$4.3\times10^{14}$ cm$^{-3}$, and the Thomas-Fermi condensate
radii are (7.6, 7.6, 0.36) $\mu$m. The $1/e$ lifetime of the
condensates in our optical trap is about 3 s.

\begin{figure}
 \includegraphics{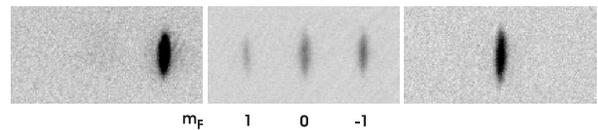}
  \caption{\label{fig:f1mag} By applying different magnetic field
  gradients at different stages of evaporation, we can create pure
  condensates of a particular spin or mixtures, which consists of
  30,000 to 15,000 atoms as seen above.
  Each absorptive image is taken after 6 ms of free expansion,
  and a weak Stern-Gerlach field is applied during the first
  2 ms of expansion to separate three spin components
  spatially. The field of view of each image is $460\times 230$ $\mu$m.}
\end{figure}

To control the initial spin population, we apply different
magnetic field gradients during the evaporation process
\cite{Chapman02}.  To create a pure $F=1,m_{F}=0$ condensate, we
apply a field gradient of 28 G/cm during the final 1 s of
evaporation. To create an equally mixed $m_{F}=-1,0,1$ condensate,
a smaller gradient (14 G/cm) is applied.  We can also create a
pure $m_{F}=-1$ by applying a 28 G/cm gradient at an earlier time
before the transfer to the lattice. Typical spinor condensates
with different spin configurations are illustrated in Figure
\ref{fig:f1mag}.

\begin{figure}
 \includegraphics{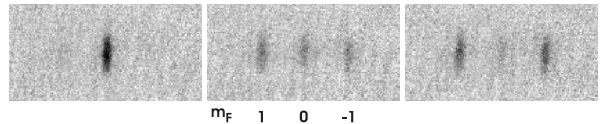}
  \caption{\label{fig:f1n0evol} Spin mixing of spinor condensate
  in the optical trap. The mixing process starts with a pure
  $m_{F}=0$ condensate in the optical trap. Three
  separate measurements of the spin state are shown after 2 s
  of spin mixing.}
\end{figure}

To study the spin mixing dynamics, we begin with pure $m_{F}=0$
condensates as the initial condition.  After condensation, the
field gradient is turned off, and a variable magnetic bias field
is adiabatically ramped up in 10 ms. This field can be directed
either along the tight (axial) or weak (radial) axes of the
confinement potential. The condensate is then allowed to evolve
for a variable amount of time, and then the spin populations are
measured using absorptive imaging following 6 ms of free
expansion.  To spatially separate the spin components, a weak
Stern-Gerlach field is applied during the first 2 ms of expansion.
Typical results for 2 s of spin mixing are shown in Figure
\ref{fig:f1n0evol}. We note from these three images taken under
identical conditions that there is significant variation on the
degree of mixing from run to run of the experiment. However, in
each case, the magnetization appears to be conserved. Generally,
the components of the spin mixtures are identical in shape to the
original cloud to within our imaging resolution.  Occasionally one
or more of the components will appear to have either a thermal
component or a distorted shape.

\begin{figure}
 \includegraphics{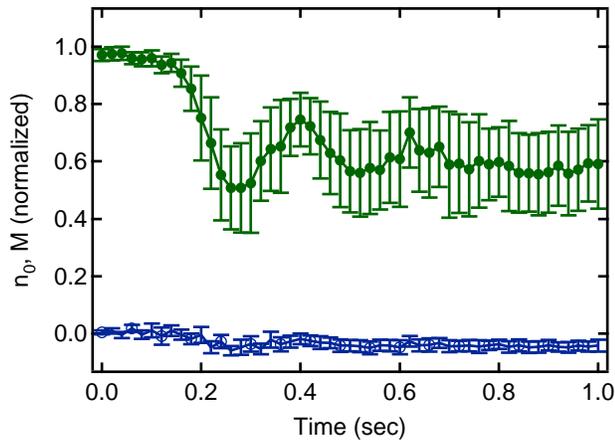}
  \caption{\label{fig:spinosc}
  Plot of the fraction in the $m_{F}$=0 state $n_{0}$
  and magnetization
  $M$ (open circle) vs.\ time. In this measurement, $M$ is
  determined to be 0.5$\pm 0.8$\% initially and -3.5$\pm 2$\%
  after mixing. Note that there is a 8.5-fold reduction on the
  fluctuation of $M$ compare to that of the total population
  (which is measured to be 17\%).  Clear population oscillations
  of $n_{0}$ are seen, and the fluctuation in $n_{0}$ is 6.6 fold
  of that of $M$. This data was taken at a bias field of 100 mG}
\end{figure}

We have measured the spin mixing for different evolution times
following preparation of the $m_{F}=0$ condensates.  The time to
reach equilibrium of the spin mixing is typically less than 600
ms, and this time decreases slightly with increasing magnetic
field. Fig.\ \ref{fig:spinosc} shows the average time evolution of
$n_{0}$ and $M$ for 50 repeated measurements. First, we note that
the magnetization, $M$, is conserved throughout the mixing to
within a few percent.  Although the data does suggest a drift of
$M$ below zero by a small amount -3.5$\pm 2$\% (the uncertainty
here and error bars in the data are purely statistical), the
deviation, if any, is comparable to our uncertainties in measuring
populations ($\sim$3\%), limited by the absorptive imaging
technique. Secondly, there is an almost 10-fold reduction in the
statistical noise of the magnetization relative to that of the
total population, which varies 17\% from shot to shot. This
suggests that the fluctuations of $n_{1}$ and $n_{-1}$, which are
coupled from the mixing processes in Eq.\ (\ref{eqn:spin-mix}),
are quantum correlated. These correlations underly theoretical
predictions for spin squeezing and entanglement in the system
\cite{You032}. Thirdly, the relaxation of $n_{0}$ to the steady
state value is not monotonic but instead shows a damped
oscillation at 4 Hz. Such oscillations are a natural outcome of
coherent spinor mixing as shown theoretically in \cite{Bigelow99}.

We attempted to directly measure the phase relationship between
the spinor components \cite{MachidaOhmi00} by performing an
interference experiment between two spinor condensates created in
adjacent lattice sites. In the absence of dephasing mechanisms, we
would expect to see clear interference fringes in time-of-flight
measurements both before and after spin mixing. In  the
measurement, clear interference fringes were visible initially,
while fringes were typically not observed after mixing. It is
quite possible however that the observed dephasing of the
interference pattern was caused by small field variations between
two sites due to a stray magnetic field gradient ($<$ 20 mG/cm)
\cite{muWave}.

\begin{figure}
 \includegraphics{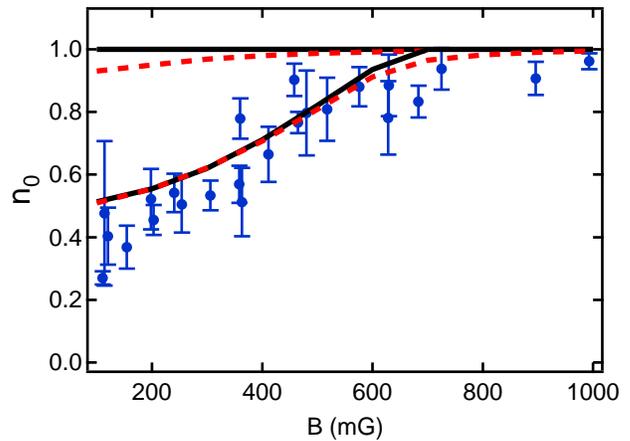}
  \caption{\label{fig:n0vsb} Plot of $n_{0}$ vs.\ magnetic
  field after 3 s of spin mixing.  The upper curves are the theoretical
  predictions for the anti-ferromagnetic case, while the lower curves
  are for the ferromagnetic case. The dashed lines are the predictions
  with a field of gradient 20 mG/cm applied, which is the measured upper bound
  in our trap. }
\end{figure}

To make comparison with theoretical predictions \cite{You031}, we
measure the degree of mixing for different applied magnetic
fields. Fig.\ \ref{fig:n0vsb} shows the results of spin mixing for
3 s, in which n$_0$ is plotted vs.\ the applied field. Also shown
in this figure are theoretical curves for both the antiferro- and
ferromagnetic cases in a magnetic field with and without a small
(20 mG/cm) field gradient  \cite{You031}. As evidenced by the
data, the spin mixing agrees with the ferromagnetic predictions
and is inconsistent with the anti-ferromagnetic prediction. When
the field is larger than 700 mG, the quadratic Zeeman effect
completely dominates the spin interaction, and the condensates
remain in the $m_{F}=0$ state. Note that magnetic fields are
applied either along tight trap or weak trap axes with different
polarities; however, no significant difference in the measurements
is observed.

We also measured spin mixing for fields less than 100 mG but found
that our results were affected by the stray $\sim$10 mG AC
magnetic fields present in the chamber.  These fields are capable
of driving off-resonant rf-transitions between the Zeeman
sub-levels. We observed this directly by creating a $F=1,m_{F}=-1$
condensate (for which there is no spin mixing due to conservation
of magnetization) and measuring the final spin population. For
magnetic fields greater than 100 mG, the magnetization remains
conserved, while at lower fields, the $m_{F}=-1$ atoms are quickly
pumped by the AC fields (within 100 ms for fields $<$10 mG) to
other Zeeman states.

\begin{figure}
(a)
\includegraphics{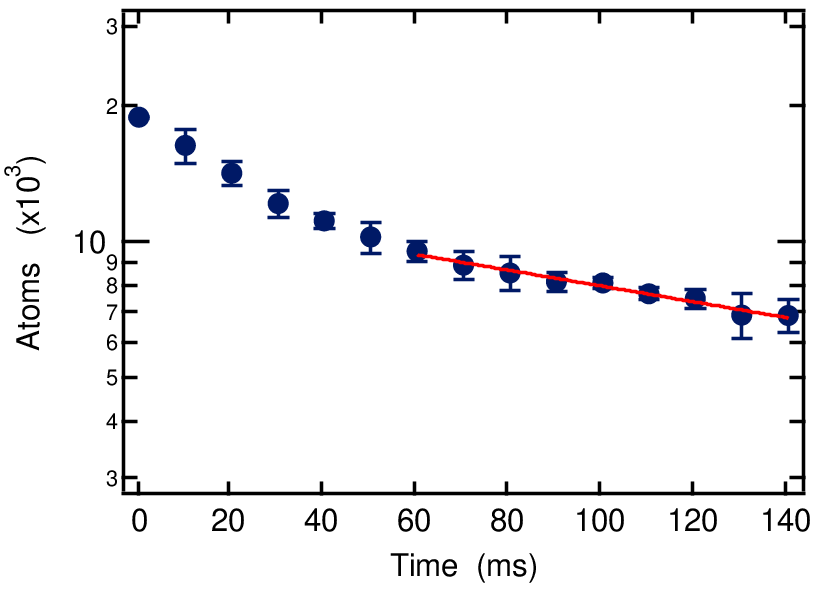}
(b)
\includegraphics{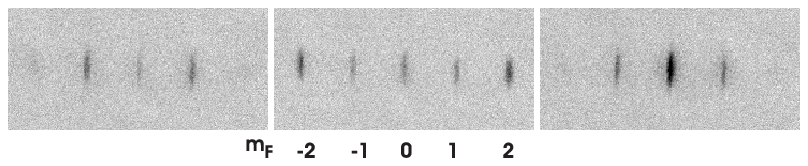}
  \caption{\label{fig:f2n0evol}
  (a) Lifetime measurement of F=2 spinor condensates. Following a rapid
  decay initial decay, the population decays exponentially with a time constant
  of 250 ms. (b) Spin mixing of the F=2 spinor condensates
  in the optical trap. These images represent three identical measurements
  after 50 ms of spin mixing. Note that the magnetization is conserved.}
\end{figure}

To study F=2 spinors, we coherently excite the pure $F=1,m_{F}=0$
condensates to $F=2,m_{F}=0$ using microwave fields tuned to 6.8
GHz.  Additionally, by controlling the bias field, the microwave
frequency, and the initial Zeeman state in the F=1 manifold, we
can pump the condensate to any Zeeman sub-level of the excited
hyperfine manifold. The F=2, m=0 condensate is observed to decay
as shown in Fig \ref{fig:f2n0evol}(a). Following an initially
rapid decay, it decays exponentially with a time constant of 250
ms. Despite the short lifetime in the excited hyperfine manifold,
we still observe spin mixing within 50 ms, and magnetization
conservation is also observed during the mixing, as shown in Fig.\
\ref{fig:f2n0evol}(b).

In summary, we have observed spin mixing of $^{87}$Rb spinor
condensates in F=1 and F=2 hyperfine manifolds in an optical trap.
The observed equilibrium spinor configurations of the lower
manifold confirms that F=1 $^{87}$Rb is ferromagnetic.  The
magnetization was conserved within the measurement errors during
the entire spin mixing process. The reduced noise of magnetization
suggests quantum correlation of the spin dynamics, which underlies
spin squeezing and spin entanglement. Future work will study the
coherence of spin mixing, spin squeezing, and entanglement.

We acknowledge valuable discussions with L.-T. Ho, T. A. B.
Kennedy, A. Kuzmich, H. Pu, C. Raman, and S. Yi.  This work is
supported by NASA Grant No. NAG3-2893. and by NSF Grant No.
PHY-0303013.


\begin{references}
\bibitem{muWave} To determine magnetic fields at trap location, we employ microwave spectroscopy on the field sensitive F=1 $\rightarrow$ F=2 hyperfine transitions which is sensitive to within 2 mG.  Furthermore, the field gradient is determined to be $<$ 20 mG/cm by measuring fields at different trap locations.
\end{references}
\end{document}